\input harvmac
\input epsf
\input amssym
%
%
\noblackbox
\newcount\figno
\figno=0
\def\fig#1#2#3{
\par\begingroup\parindent=0pt\leftskip=1cm\rightskip=1cm\parindent=0pt
\baselineskip=11pt
\global\advance\figno by 1
\midinsert
\epsfxsize=#3
\centerline{\epsfbox{#2}}
\vskip -21pt
{\bf Fig.\ \the\figno: } #1\par
\endinsert\endgroup\par
}
\def\figlabel#1{\xdef#1{\the\figno}}
\def\encadremath#1{\vbox{\hrule\hbox{\vrule\kern8pt\vbox{\kern8pt
\hbox{$\displaystyle #1$}\kern8pt}
\kern8pt\vrule}\hrule}}

\def\p{\partial}

\def\At{\tilde{A}}
\def\del{\nabla}
\def\eps{\epsilon}

\def\At{\tilde{A}}

\def\eps{\epsilon}

\def\Aa{A^{(0)}}
\def\Ab{A^{(1)}}

\def\ba{b^{(1)}}

\def\Bc{{\cal B}}

\def\ve{\varepsilon}


\Title{\vbox{\baselineskip12pt
}} {\vbox{\centerline {S-duality in AdS/CFT magnetohydrodynamics}}} \centerline{James
Hansen\foot{jhansen@physics.ucla.edu} and Per
Kraus\foot{pkraus@ucla.edu}}

\bigskip
\centerline{${}$\it{Department of Physics and Astronomy,
UCLA,}}\centerline{\it{ Los Angeles, CA 90095-1547, USA.}}

\baselineskip15pt

\vskip .3in

\centerline{\bf Abstract}

We study the nonlinear hydrodynamics of a 2+1 dimensional charged conformal fluid subject to slowly varying external electric and magnetic  fields.  Following recent work on deriving  nonlinear hydrodynamics from gravity, we demonstrate how long wavelength perturbations of the AdS dyonic black brane  solution of 4D supergravity are governed by equations equivalent to fluid dynamics equations in the boundary theory.  We investigate the implications of $S$-duality for our system, and derive restrictions imposed on the transport coefficients of a generic fluid invariant under the $S$ operation.  We also expand on our earlier work and determine a new set of previously undetermined transport coefficients for the conformal fluid with an AdS gravity dual.  Quite surprisingly, we discover that half of the transport coefficients allowed by symmetry vanish in the holographic fluid at linear order in the hydrodynamic expansion.

\Date{July, 2009}
\baselineskip14pt


\lref\HorowitzJD{
  G.~T.~Horowitz and V.~E.~Hubeny,
  ``Quasinormal modes of AdS black holes and the approach to thermal
  equilibrium,''
  Phys.\ Rev.\  D {\bf 62}, 024027 (2000)
  [arXiv:hep-th/9909056].
}

\lref\GaillardRJ{
  M.~K.~Gaillard and B.~Zumino,
  ``Duality Rotations For Interacting Fields,''
  Nucl.\ Phys.\  B {\bf 193}, 221 (1981).
}

\lref\WittenYA{
  E.~Witten,
  ``SL(2,Z) action on three-dimensional conformal field theories with Abelian
  symmetry,''
  arXiv:hep-th/0307041.
}

\lref\NMFG{
  J.~Hansen and P.~Kraus,
  ``Nonlinear Magnetohydrodynamics from Gravity,''
  JHEP {\bf 0904}, 048 (2009)
  [arXiv:0811.3468 [hep-th]].
}

\lref\Hansen{
  J.~Hansen and P.~Kraus,
  JHEP {\bf 0612}, 009 (2006)
  [arXiv:hep-th/0606230].
}

\lref\SonSD{
  D.~T.~Son and A.~O.~Starinets,
  JHEP {\bf 0209}, 042 (2002)
  [arXiv:hep-th/0205051].
}

\lref\SonVK{
  D.~T.~Son and A.~O.~Starinets,
  ``Viscosity, Black Holes, and Quantum Field Theory,''
  Ann.\ Rev.\ Nucl.\ Part.\ Sci.\  {\bf 57}, 95 (2007)
  [arXiv:0704.0240 [hep-th]].
}

\lref\CFTQNN{
  D.~Birmingham, I.~Sachs and S.~N.~Solodukhin,
  ``Conformal field theory interpretation of black hole quasi-normal 
modes,''
  Phys.\ Rev.\ Lett.\  {\bf 88}, 151301 (2002)
  [arXiv:hep-th/0112055].
}

\lref\SK{
  C.~P.~Herzog and D.~T.~Son,
  ``Schwinger-Keldysh propagators from AdS/CFT correspondence,''
  JHEP {\bf 0303}, 046 (2003)
  [arXiv:hep-th/0212072].
}

\lref\HaackXX{
  M.~Haack and A.~Yarom,
  ``Universality of second order transport coefficients from the gauge-string
  duality,''
  arXiv:0811.1794 [hep-th].
}

\lref\NLFDFG{
  S.~Bhattacharyya, V.~E.~Hubeny, S.~Minwalla and M.~Rangamani,
  ``Nonlinear Fluid Dynamics from Gravity,''
  JHEP {\bf 0802}, 045 (2008)
  [arXiv:0712.2456 [hep-th]].
}

\lref\CJ{
  A.~Chamblin, R.~Emparan, C.~V.~Johnson and R.~C.~Myers,
  ``Charged AdS black holes and catastrophic holography,''
  Phys.\ Rev.\  D {\bf 60}, 064018 (1999)
  [arXiv:hep-th/9902170].
}

\lref\SH{
  P.~Kovtun, D.~T.~Son and A.~O.~Starinets,
  ``Holography and hydrodynamics: Diffusion on stretched horizons,''
  JHEP {\bf 0310}, 064 (2003)
  [arXiv:hep-th/0309213].
}

\lref\KovtunKX{
  P.~Kovtun and A.~Ritz,
  ``Universal conductivity and central charges,''
  arXiv:0806.0110 [hep-th].
}

\lref\VanRaamsdonkFP{
  M.~Van Raamsdonk,
  ``Black Hole Dynamics From Atmospheric Science,''
  JHEP {\bf 0805}, 106 (2008)
  [arXiv:0802.3224 [hep-th]].
}

\lref\BhattacharyyaXC{
  S.~Bhattacharyya {\it et al.},
  ``Local Fluid Dynamical Entropy from Gravity,''
  JHEP {\bf 0806}, 055 (2008)
  [arXiv:0803.2526 [hep-th]].
}

\lref\BhattacharyyaJI{
  S.~Bhattacharyya, R.~Loganayagam, S.~Minwalla, S.~Nampuri, S.~P.~Trivedi and S.~R.~Wadia,
  ``Forced Fluid Dynamics from Gravity,''
  arXiv:0806.0006 [hep-th].
}

\lref\BhattacharyyaMZ{
  S.~Bhattacharyya, R.~Loganayagam, I.~Mandal, S.~Minwalla and A.~Sharma,
  ``Conformal Nonlinear Fluid Dynamics from Gravity in Arbitrary Dimensions,''
  arXiv:0809.4272 [hep-th].
}

\lref\BanerjeeTH{
  N.~Banerjee, J.~Bhattacharya, S.~Bhattacharyya, S.~Dutta, R.~Loganayagam and P.~Surowka,
  ``Hydrodynamics from charged black branes,''
  arXiv:0809.2596 [hep-th].
}

\lref\HaackCP{
  M.~Haack and A.~Yarom,
  ``Nonlinear viscous hydrodynamics in various dimensions using AdS/CFT,''
  JHEP {\bf 0810}, 063 (2008)
  [arXiv:0806.4602 [hep-th]].
}

\lref\ErdmengerRM{
  J.~Erdmenger, M.~Haack, M.~Kaminski and A.~Yarom,
  ``Fluid dynamics of R-charged black holes,''
  arXiv:0809.2488 [hep-th].
}

\lref\HartnollIP{
  S.~A.~Hartnoll and C.~P.~Herzog,
  ``Ohm's Law at strong coupling: S duality and the cyclotron resonance,''
  Phys.\ Rev.\  D {\bf 76}, 106012 (2007)
  [arXiv:0706.3228 [hep-th]].
}

\lref\HartnollIH{
  S.~A.~Hartnoll, P.~K.~Kovtun, M.~Muller and S.~Sachdev,
  ``Theory of the Nernst effect near quantum phase transitions in condensed
  matter, and in dyonic black holes,''
  Phys.\ Rev.\  B {\bf 76}, 144502 (2007)
  [arXiv:0706.3215 [cond-mat.str-el]].
}

\lref\HartnollAI{
  S.~A.~Hartnoll and P.~Kovtun,
  ``Hall conductivity from dyonic black holes,''
  Phys.\ Rev.\  D {\bf 76}, 066001 (2007)
  [arXiv:0704.1160 [hep-th]].
}

\lref\HerzogIJ{
  C.~P.~Herzog, P.~Kovtun, S.~Sachdev and D.~T.~Son,
  ``Quantum critical transport, duality, and M-theory,''
  Phys.\ Rev.\  D {\bf 75}, 085020 (2007)
  [arXiv:hep-th/0701036].
}

\lref\BalasubramanianRE{
  V.~Balasubramanian and P.~Kraus,
  ``A stress tensor for anti-de Sitter gravity,''
  Commun.\ Math.\ Phys.\  {\bf 208}, 413 (1999)
  [arXiv:hep-th/9902121].
}

\lref\HenningsonGX{
  M.~Henningson and K.~Skenderis,
  ``The holographic Weyl anomaly,''
  JHEP {\bf 9807}, 023 (1998)
  [arXiv:hep-th/9806087].
}

\lref\BuchbinderDC{
  E.~I.~Buchbinder, A.~Buchel and S.~E.~Vazquez,
  ``Sound Waves in (2+1) Dimensional Holographic Magnetic Fluids,''
  arXiv:0810.4094 [hep-th].
}

\lref\BuchbinderNF{
  E.~I.~Buchbinder and A.~Buchel,
  arXiv:0811.4325 [hep-th].
}

\lref\RangamaniXK{
  M.~Rangamani,
  ``Gravity \& Hydrodynamics: Lectures on the fluid-gravity correspondence,''
  arXiv:0905.4352 [hep-th].
}

\newsec{Introduction}

An interesting recent development has been the application of AdS/CFT to the nonlinear hydrodynamics of strongly coupled field theories \refs{\NLFDFG}  (some early applications include \refs{\VanRaamsdonkFP,\BhattacharyyaXC,\BhattacharyyaJI,
\BhattacharyyaMZ,\BanerjeeTH,\HaackCP,\ErdmengerRM}; see also the review \RangamaniXK\ for further references, as well as the review \SonVK\ for earlier work focussed on the linearized regime.)   One system of interest,
relevant to various phenomena including  superconductivity, graphene, and the quantum Hall effect, is that of a charged 2+1 dimensional conformal fluid evolving under the influence of  external electromagnetic fields.  In a previous paper \NMFG\  we applied the method of \NLFDFG\ to calculate a subset of linear and nonlinear transport coefficients in fluids with an Einstein-Maxwell gravity dual.  In this paper we extend these results to a complete calculation of all the transport coefficients in this sector up to second order in the hydrodynamic expansion.   The main generalization involves allowing the external electromagnetic field to vary in space and time.\foot{A natural further generalization
is to allow the fluid to live on a curved  geometry, as in \BhattacharyyaJI.    This generalization is discussed in section 2 to first order.}
Along the way, we study the role of $S$-duality in these systems and discover a surprising cancelation in the linear order transport coefficients.

It was noted in \WittenYA\ that there is a natural $SL(2,Z)$ action on the space of 2+1 dimensional conformal field theories with a $U(1)$ global symmetry. Generically this action is a duality: it maps solutions of one CFT to solutions of some other inequivalent CFT.  However, in some cases it makes sense to consider CFTs invariant under  the $S$ element of $SL(2,Z)$.  Such CFTs arise naturally in the AdS/CFT context:  if a CFT has a gravity dual that can be consistently truncated to Einstein-Maxwell gravity, then the boundary $S$ operation is mapped to electric-magnetic duality in the bulk, which is a symmetry of the Einstein-Maxwell equations of motion.  The implications  for AdS/CFT linear transport properties were discussed extensively in \refs{\HerzogIJ,\HartnollIP}.   In section 2, we will be interested in studying the significance of this invariance for generic $S$-invariant CFT hydrodynamics.  We show how $S$ invariance tightly constrains the hydrodynamics of a generic CFT, so that in general, the complete fluid  equations of motion at leading order in derivatives are completely specified by the  equation of state together with one non-negative real function and one real number. This is a general statement about
S-invariant conformal fluids, independent of the AdS/CFT correspondence.

In sections 3 and 4, we use the dyonic black brane solution of $3+1$ dimensional Einstein-Maxwell gravity to construct a gravity dual for hydrodynamic fluctuations.    Following \NLFDFG\ we proceed order by order in a derivative expansion to arrive at a solution in  local but not global thermodynamic equilibrium.  The starting
point is a configuration in global equilibrium, with free parameters corresponding to  energy and charge densities along with the value of a constant background magnetic field.  We then
allow these parameters to vary slowly in spacetime, and also allow for a slowly varying electric field.   Associated with these varying parameters are numerous transport coefficients expressing the flow of current and stress-energy.  By solving the Einstein-Maxwell equations in the boundary derivative expansion,   we determine all these transport coefficients up to second order in derivatives.

As in our previous work we consider the case of {\it magnetohydrodynamics}, such that
the external magnetic field is nonzero in any Lorentz frame; that is, we assume $B^2 > \vec{E}^2$ at all points in the fluid.  Besides the length scale associated with the
temperature, $l_T \sim {1\over T}$, there is then a second length scale set by the
magnetic field, $l_B \sim {1\over \sqrt{B}}$, where $B$ denotes the value of the magnetic field in a Lorentz frame with $\vec{E}=0$.   The  fluid degrees of freedom
are assumed to be slowly varying over both of these length scales.  Our approach is to be distinguished from that of \refs{\HartnollIH,\HartnollIP} (see also \refs{\BuchbinderDC,\BuchbinderNF}), which is more appropriate for weak $B$ fields (and is also restricted to the linear regime).

One result of our calculations is that we discover an interesting cancelation in the linear transport coefficients for the fluid dual to Einstein-Maxwell gravity.  These transport coefficients are naturally grouped  into two sectors according to their   behavior under the $S$ operation:  the energy diffusion and electric conductivity sector, which was studied in \refs{\HartnollIH,\HartnollIP}; and a sector describing charge diffusion and the response to variations in the magnetic field.  In each of these sectors we find that half of the transport coefficients vanish when written in the natural S-covariant basis.   This is surprising, inasmuch as
we demonstrate that nonvanishing values are consistent with all the symmetries and with
positive divergence of the entropy current.  The vanishing results  therefore seem
particular to fluids with an Einstein-Maxwell dual description.  Of course, it would be
especially interesting to find a real world fluid with such vanishing --- or at least small ---
transport coefficients, as such a fluid would provide a very promising system for making
experimental contact with AdS/CFT.

\newsec{Hydrodynamics of a general $S$-invariant conformal fluid}

We are interested in studying the hydrodynamics of a 2+1 dimensional CFT with a conserved current  coupled to  slowly varying external electric and magnetic fields.  We consider the case where the external field is locally magnetic; that is we impose a constraint that $|\vec{E}| < |B|$ everywhere so that there will always exist a boosted frame at each point where $\vec{E} = 0$.  Our fluid dynamical variables are taken to be the energy density, $T^{00} =\epsilon$, and the charge density, $J^0 =\rho$.  In particular, we do not take the fluid velocity to be an independent degree of freedom; the reason for this, as discussed extensively in  \NMFG, is that in the presence of a background magnetic field momentum is not a conserved quantity. At large $B$ and in the fluid dynamical derivative expansion, the fluid momentum density is fixed by the equations of motion in terms of $\epsilon$ and $\rho$, and so introducing additional degrees of freedom to represent it would be a needless complication.

The equations of motion for our fluid are the conservation equations $\p_\nu T^{\mu\nu} = F^{\mu \nu}J_\nu$ and $\p_\mu J^\mu =0$.   We solve the equations
by working in the neighborhood of an arbitrary point in the fluid, and
choose a Lorentz frame such that $\vec{E}=0$ at that point.   Away from the chosen point
the electric field will be nonzero but small, in the sense that it will scale like a
derivative in our fluid dynamical derivative expansion; in particular, its magnitude will be of the same hydrodynamic order as derivatives of $\epsilon$, $\rho$ and $B$.
In this frame it is natural to write out the equations of motion as
\eqn\Aa{\eqalign{\p_\mu T^{\mu 0}&= -E_i J^i~,\cr  \p_\mu T^{\mu i}&= -\rho E^i +B \epsilon^{ij}J_j~,\cr \p_0 \rho&=- \p_i J^i~,\cr  \p_0 B& = \epsilon^{ij} \p_i E_j~,  }}
with $F_{0i}=E_i$,   $F_{ij}= B \epsilon_{ij}$, and in the last line we have written out the Bianchi identity.  In this equation we have introduced a notation that we will use throughout, where lowercase roman indices  refer to the two spatial directions in our fluid.  In terms of the quantities in \Aa, the $SL(2, Z)$ duality of \WittenYA\ corresponds to the following $S$ operation \foot{In the appendix we show how the $SL(2,Z)$ of a boundary CFT can be holographically related to the $SL(2,Z)$ of electric-magnetic duality in the bulk for CFTs with a gravity dual.} :
\eqn\Ab{\eqalign{ \rho &\rightarrow B \cr
B &\rightarrow -\rho \cr
E_i  &\rightarrow -\epsilon_{ij}J^j \cr
J^i &\rightarrow -\epsilon^{ij}E_j ~, }}
as well as the $T$ operation:
\eqn\Abc{\eqalign{ \rho &\rightarrow \rho +B \cr
B &\rightarrow B \cr
E_i  &\rightarrow E_i \cr
J^i &\rightarrow J^i -\epsilon^{ij}E_j ~.  }}
The $S$ operation can be viewed as  particle-vortex duality, in that it interchanges a unit of electric charge with a unit of magnetic flux.
The $S$ and $T$ operations are symmetries of the equations of motion \Aa\ provided that the
stress tensor is invariant.
Repeated application of $S$ and $T$ generates the full $SL(2,Z)$, acting as
\eqn\Ac{ \left(\matrix{\rho \cr B}\right) \rightarrow  \left(\matrix{a&b \cr c& d}\right)\left(\matrix{\rho \cr B}\right)~,\quad\quad \left(\matrix{E_i  \cr -\epsilon_{ij}J^j }\right) \rightarrow  \left(\matrix{a&b \cr c& d}\right)\left(\matrix{E_i  \cr -\epsilon_{ij}J^j }\right)~,}
with $ad-bc=1$.

In general, $S$ and $T$ map solutions of one CFT to solutions of a different CFT. The equations of fluid dynamics are specified not only by \Aa\ but also by the constitutive relations which express the stress tensor and charge current in terms of the fluid variables. These relations may or may not be invariant under  $SL(2,Z)$.  However, it is not hard to see that in theories for which the stress  tensor is invariant under $S$, the full equations of motion of our hydrodynamic theory will be invariant as well.  In fact, as noted in \WittenYA, and as we will see in section 3, conformal theories with a gravity dual will be $S$ invariant in sectors dual to Einstein-Maxwell gravity.  It is these CFTs which we will be considering in the remainder of this paper.

Unlike  $S$, the $T$ of $SL(2,Z)$ will not appear as a symmetry of our equations of motion.
To appreciate the distinction, note that the action of $T$ is the same as the redefinition of charge and current that results from  adding a Chern-Simons term to the action of our CFT.  If we think of theories as being specified by their action, then the $T$ operation relates distinct theories.
By setting the coefficient of the Chern-Simons term to some fixed value (we will take it to be zero), there is then no constraint on the constitutive relations that they
be invariant under $T$.  By contrast, the $S$ operation relates the theory to itself,
and we will demand invariance under $S$.  From the bulk gravity point of view, $T$
corresponds to shifting the $\theta$-angle, while $S$ corresponds to electric-magnetic
duality  \WittenYA.

To specify the equations of motion for a general $S$ invariant fluid we need to express the
stress tensor in terms of an $S$ invariant derivative expansions.  It turns out that the combinations
\eqn\Ad{\eqalign{
M_{ij}^{\pm} = \rho \delta_{ij} \mp B \eps_{ij} \cr
N_{i}^{\pm} = E_i \mp J_i~,
}}
will be particularly useful.  These quantities transform under $S$ as:
\eqn\Ae{\eqalign{
M_{ij}^{\pm} \rightarrow \pm \eps_{ik} M_{kj}^{\pm} \cr
N_i^{\pm} \rightarrow \pm \eps_{ij} N_j^{\pm}.
}}
These objects are even under parity, and odd under charge conjugation $C$.  Time reversal interchanges the $+$ and $-$ representations.  Because the currents and electric field both vanish in the hydrodynamic limit, $M_{ij}^{\pm}$ is zeroth order in our derivative expansion while $N_i^{\pm}$ is order one.

There are only two zeroth order $S$-invariant scalars in this theory, the energy density $\eps$ in the energy-electric field sector and the quantity
\eqn\Af{ M = {1\over 2 } M^+_{ij}M^-_{ji} = \rho^2 +B^2~,}
in the charge-magnetic field sector.

\subsec{Constitutive equations and transport coefficients}

To completely specify our fluid dynamics we just need to write down constitutive relations for $T^{\mu \nu}$ and $J^{\mu}$ in terms of fluid variables $\epsilon$  and $\rho$, and the background fields $B$ and $E_i$.\foot{We take the fluid variables to be $\epsilon$ and $\rho$ rather than their thermodynamic conjugates $T$ and $\mu$.  This is appropriate, since it is only the former that have an unambiguous meaning away from thermal equilibrium.   For small deviations from equilibrium one can of course go back and forth between the two using the equation of state.}   This is carried out order by order in a derivative expansion, where $\epsilon$, $\rho$, $B$, and $E_i$ are all allowed to slowly vary.  At zeroth order in derivatives we just have the equilibrium fluid, and we can
assume without loss of generality that $E_i=0$ at lowest order in derivatives by locally boosting to an appropriate frame.   The equilibrium fluid is thus labeled by $(\epsilon, \rho, B)$.

Given our assumption that $B$ is nonzero, we do not actually need to provide a constitutive
relation for the current.   This follows since we can solve for $J^i$ in the second line of \Aa\ as
\eqn\Aga{
J_i = - {1\over B} \eps_{ij}  \left(\p_{\mu} T^{\mu j} + \rho E^j \right) ~.
}
The current is thus  completely determined in terms of other quantities.   Note that the
choice to solve for $J_i$ is not invariant under $S$;  the S-dual choice would be to
solve for $E_i$, which is allowed provided that $\rho$ is nonzero.  The point is that
although our fluid  equations of motion  are assumed to be S-invariant, we are
choosing to solve them in a non-manifestly S-invariant manner.   This asymmetry
under S-duality is built in from the assumption that $B$ and $E_i$ are regarded as fixed
external fields, while $\rho$ and $J^i$ are regarded as dynamical variables, even though
these two sets of quantities are exchanged under S-duality.

The form of the stress tensor is constrained by symmetries. We demand invariance under charge conjugation, under which $M^\pm_{ij}$ and $N^\pm_i$ are both odd.  Also, we require invariance under spatial $SO(2)$ rotations and parity; this requires that $i$ type indices match up on
both sides of any equation.   Note that we do not demand invariance under time reversal since, by definition, dissipative fluid dynamics has a preferred direction of time.  Finally, we impose the scale invariance of our theory, which implies tracelessness of the stress tensor, and hence fixes $T^{ij} = {1\over 2}\epsilon \delta_{ij}$ at zeroth order in derivatives.  Because $T^{00} = \eps$ at all orders in the derivative expansion, this leaves only $T^{0i}$ and the symmetric traceless part of $T^{ij}$ to be determined.

At first order in the derivative expansion we can define the following four vectors sharing
the same symmetries as $T^{0i}$:
\eqn\Ag{\eqalign{
x^1_i &= M_{ij}^+ N_j^- \cr
x^2_i &= M_{ij}^- N_j^+ \cr
x^3_i &= M_{ij}^+ \p_k M_{jk}^- \cr
x^4_i &= M_{ij}^- \p_k M_{jk}^+ ~.
}}
The symmetries then fix $T^{0i}$ to be of the form
\eqn\Ah{ T^{0i} = c_1 x_i^1 +c_2 x^2_i + c_3 x_i^3 + c_4 x_i^4~,}
where $c_a=c_a(\epsilon,M)$ are, by definition,  transport coefficients.    We can use dimensional analysis to express each of the coefficient functions $c_a$ as
a power of $\epsilon$ times an arbitrary function of the dimensionless combination
$\epsilon^4/M^3$.   In the formula \Ah\ $T^{0i}$ appears to be a function of the current  $J^i$; however, it is implicit that $J^i$ at first order is reexpressed in terms of first derivatives of $(\epsilon,\rho)$ through \Aga. Hence $T^{0i}$ is really a function of the fluid variables $(\epsilon,\rho)$ and the background fields $(B,E_i)$.  There are no contributions to $T^{ij}$ at this order in derivatives.

\subsec{Entropy current}

Studying fluctuations of the  dyonic black brane solution will lead to specific functions $c_a$, along with
their analogs at second order.  But before turning to that computation it is useful
to consider the general constraints on these functions.  A set of fluid equations is only
physically permissible if it is possible to define an entropy current $S^\mu$ whose divergence is positive semi-definite, $\p_\mu S^\mu \geq 0$.   This expresses the condition that locally entropy should  be  produced and not destroyed.  By definition we have
$S^0=s$, where $s=s(\eps,M)$ is the entropy density of the equilibrium fluid.   To establish consistency we then need find the  spatial components $S^i$ compatible with positive
divergence; existence of these spatial components will be seen to imply constraints on the transport coefficients $c_a$.  As with $T^{0i}$, the symmetries constrain $S^i$ at first order in derivatives to be of the form
\eqn\Ai{S^i =  \alpha_1 x_i^1 +\alpha_2 x^2_i + \alpha_3 x_i^3 + \alpha_4 x_i^4~,}
with $\alpha_a = \alpha_a(\epsilon,M)$ constrained by dimensional analysis.

Using the equations of motion, we can compute $\p_\mu S^\mu$ in terms of second order quantities and squares of first order quantities.   Positivity of the divergence for arbitrary configurations requires that all the second order linear contributions vanish, and that the remaining terms form a sum of squares with positive coefficients.

Computing the terms linear in second order quantities we find
\eqn\Aj{\eqalign{
\left[ \p_{\mu} S^\mu \right]_{{\rm linear}} &= \bigg(\alpha_1-\p_\eps s c_1 -\p_M s\bigg) M_{ij}^+ \p_i N_j^-  +  \bigg(\alpha_2-\p_\eps s c_2 +\p_M s\bigg) M_{ij}^- \p_i N_j^+ \cr & +\big(\alpha_3 + \alpha_4 - \p_\eps s c_3 - \p_\eps s c_4\big) M_{ij}^+ \p_i \p_k M^-_{jk}~.}}
Setting this to zero fixes three combinations of the four coefficients $\alpha_a$.  Using the parameter $\chi$ to label the family of solutions, we write
\eqn\Ak{\eqalign{
\alpha_1 &= \p_\eps s  c_1 + \p_M s  \cr
\alpha_2 &= \p_\eps s c_2 - \p_M s \cr
\alpha_3 &= \p_\eps s c_3 + \chi \cr
\alpha_4 &= \p_\eps s c_4 - \chi ~.
}}

With this result in hand, we can proceed to compute the terms quadratic in first order
quantities.  After considerable algebra we find
\eqn\Al{\eqalign{
\p_{\mu} S^{\mu} &= \left[- c_1 \p_{\eps}^2 s  - \p_M \p_{\eps} s  - {\p_\eps s\over 4 M} \right]x_i^1 x_i^1 \cr
& + \left[- c_2 \p_{\eps}^2 s + \p_M \p_{\eps} s + {\p_\eps s\over 4 M} \right]x_i^2 x_i^2 \cr
& + \left[c_3 \p_M \p_{\eps} s + {\chi \over M} + \p_M \chi\right]x_i^3 x_i^3 \cr
& + \left[ c_4 \p_M \p_{\eps} s - {\chi \over M} - \p_M \chi \right]x_i^4 x_i^4 \cr
& + \left[ - (c_1 + c_2) \p^2_{\eps} s\right]x_i^1 x_i^2 \cr
& + \left[- c_3 \p^2_{\eps}s + \p^2_M s - \p_{\eps} \chi + c_1 \p_M \p_{\eps} s + {\p_M  s \over M} \right]x_i^1 x_i^3 \cr
& + \left[ - c_4 \p^2_{\eps} s + \p_M^2 s + \p_{\eps} \chi + c_1 \p_M \p_{\eps} s\right]x_i^1 x_i^4 \cr
& + \left[ - c_3 \p^2_{\eps} s - \p^2_M s - \p_{\eps} \chi + c_2 \p_M \p_{\eps} s \right]x_i^2 x_i^3 \cr
& + \left[ - c_4 \p^2_{\eps} s - \p^2_M s + \p_{\eps} \chi + c_2 \p_M \p_{\eps} s - {\p_M s \over M} \right]x_i^2 x_i^4 \cr
& + \left[ (c_3 + c_4) \p_M \p_{\eps} s \right]x_i^3 x_i^4~.
}}
Non-negativity of the  divergence of the entropy current now reduces to the condition that this quadratic form, viewed as a symmetric matrix, have no negative eigenvalues.

We can examine this condition separately in various subspaces.  In the $x^1_i - x^2_i$ we find the condition
\eqn\Am{ c_1-c_2 =-{1\over \p_\eps^2 s}\big(2 \p_M \p_\eps s +{\p_\eps s \over 2M}\big)~.}

Next we consider  the $x^3_i - x^4_i$ and $(x^1_i+x^2_i)-( x^3_i+x^4_i)$ subspaces, which leads to the two conditions
\eqn\Ama{\eqalign{
\p_M (M \chi) &=  - {1 \over 2} (c_3 - c_4) M\p_M \p_{\eps} s \cr
\p_{\eps} (M \chi) & = {1 \over 2}\p_M s - {1 \over 2} (c_3 - c_4)M \p^2_{\eps} s ~.
}}
Equating mixed partials gives an integrability condition
\eqn\Ao{ \bigg[ \p_\eps^2 s \p_M - \p_M \p_\eps s \p_\eps \bigg]\bigg[ M(c_3-c_4)\bigg]=
\p_M^2 s~.}
This equation determines $M(c_3-c_4)$ up to the addition of an arbitrary function of
the form $f(\p_\eps s)$, which by dimensional analysis takes the form $f(\p_\eps s) = c/(\p_\eps s)^2$. Then,  $\Am$  fixes $M\chi$ in terms of $c_3-c_4$ up to an additive constant.

Using these results the divergence becomes
\eqn\Ap{
\p_{\mu}S^{\mu} = -{1\over 2}\left[  \p_{\eps}^2 s (c_1 + c_2) \right] t^2 + {1\over 2} \left[ \p_M \p_{\eps} s(c_1 + c_2) -  \p_{\eps}^2 s (c_3 + c_4) \right] \, ut  - {1\over 2}\left[ \p_M \p_{\eps} s  (c_3 + c_4) \right] u^2~,
}
where we are writing $x_i^1 + x_i^2 = t$ and $x_i^3 + x_i^4 = u$.
The determinant of the associated matrix is
\eqn\Aq{
- {1 \over 16} \bigg( \p_{\eps}^2 s(c_3 + c_4) +  \p_M \p_{\eps} s (c_1 + c_2)  \bigg)^2
~.}
Since this is negative semi-definite we must demand that it actually vanishes, which
tells us that
\eqn\Ar{
c_3 + c_4 = -{\p_M \p_{\eps} s \over \p_{\eps}^2 s} (c_1 + c_2) ~.
}
The divergence then finally takes the form:
\eqn\As{
\p_{\mu} S^{\mu} = - {c_1 + c_2 \over \p_{\eps}^2 s} \bigg((\p_{\eps}^2 s)^2 + (\p_M \p_{\eps} s)^2  \bigg)\left[- \p_{\eps}^2 s  {x_i^1 + x_i^2 \over \p_M \p_{\eps} s} +  (x_i^3 + x_i^4) \right]^2~.
}
Since ${1\over \p_\eps^2 s }= -T^2 C$, where $C$ is the specific heat, we see that $\p_\eps^2 s <0$ and hence $ \p_{\mu} S^{\mu} \geq 0$ provided that $c_1 + c_2$ is non-negative.

Let us summarize the constraints imposed on an acceptable fluid dynamics.  $c_1+c_2$ is allowed to be an arbitrary, non-negative, dimensionally correct, function of $(\eps,M)$.   $c_1 -c_2$ is fixed
by \Am;  $c_3 + c_4$ is fixed by \Ar; and $c_3 - c_4$ is fixed by \Ao\ up to the addition of
${c\over M (\p_\eps s)^2}$, for some number $c$.    Altogether then, given an equation of state there is one free real function and one free real number labeling the space of allowed fluid equations of motion.

In the above, we considered a fluid living on a flat three dimensional spacetime, but one
might wonder whether there are any additional constraints to be found by putting the fluid
on a curved geometry.  It is easy to see that no such additional constraints can arise.
At this order in the derivative expansion, one can simply repeat the previous
computation with indices suitably contracted.  The one place where a new constraint would
arise is if a term proportional to the Ricci scalar were to appear in the expression for
$\nabla_\mu S^\mu$.    Since the Ricci scalar is not positive definite, positivity of the  divergence would impose the constraint that its coefficient be zero.   However,  this coefficient is
automatically zero.  This can be seen by considering the special case of a fluid on
$S^2 \times R$.  In this case one can obviously find static fluid configurations with
constant $\epsilon$ and $\rho$, and these configuration will obviously obey $\nabla_\mu S^\mu=0$.  But evaluated on such  a configuration, all possible terms at second order in
derivatives that can appear in the expression for $\nabla_\mu S^\mu$ vanish, except for
the Ricci scalar.  So the fact that  $\nabla_\mu S^\mu=0$ for such a configuration implies that the
coefficient of the Ricci scalar is zero.
We  conclude that our fluid equations can be  generalized to curved
spacetime with no new constraints being required.

\subsec{Black brane equation of state}

In the preceding discussion the equilibrium equation of state of the fluid is left
unspecified.  We now consider the equation of state corresponding to the dyonic black
brane considered in this paper.   The entropy density is given by $s=\pi a^2$ where
the horizon radius $a$ is given by the largest real root of
\eqn\At{ a^2 +{\rho^2 +B^2 \over a^2}-{2\eps \over a}=0~.}
If we use dimensional analysis to write
\eqn\Au{ c_1+c_2 = {1\over a} F\big({a^4/M}\big)}
and introduce the numerical constant $c$, then we find
\eqn\Av{\eqalign{ c_1 -c_2 & = {3(a^4+M)\over 4aM}\cr    c_3 +c_4&=  {(3a^4+M)\over 3a^2(a^4+M)}F\big({a^4/M}\big)\cr
c_3-c_4 &= {(3a^4+M)\over 4a^2M} +{(3a^4-M)^2 \over a^6 M} ~c~. }}
Any choice of function $F$ and constant $c$ leads to an acceptable fluid dynamics.
As we'll see later, gravity picks out one particular choice,
\eqn\Aw{  F\big({a^4/M}\big) = {3\over 4} \bigg(1+{a^4\over M} \bigg)~,\quad\quad c=0~.}
corresponding to the transport coefficients
\eqn\Ax{\eqalign{c_1 & = {3(a^4+M) \over 4aM} \cr
c_2 &=0 \cr
c_3 & = {3a^4 +M \over 4a^2 M}
\cr
c_4 &=0~.}}
It is interesting that gravity yields this simple result in which two of the naturally
defined transport coefficients vanish. It would be interesting to know if there is some
general reason for this to be the case since,  as emphasized above, all the general  consistency requirements on the fluid equations can be satisfied even when these coefficients are non-vanishing.

\newsec{The gravity dual description}

We now turn to our calculation of the transport coefficients for our fluid with an AdS gravity dual description.  As developed in \refs{\HartnollAI,\HartnollIH}, the dyonic black brane solution to 4-dimensional Einstein-Maxwell gravity provides a dual description of a finite temperature 2+1 dimensional CFT  at nonzero charge density and under the influence of an external magnetic field. In this section, we first review the  duality dictionary in the case of global thermal equilibrium, and then proceed to develop a derivative expansion that will allow us to study the duality in the hydrodynamic regime.

\subsec{The dyonic black brane}

The 4-dimensional bulk Einstein-Maxwell action is given by
\eqn\Ba{
S= {2 \over \kappa_4^2} \int d^4x \sqrt{-g} \left[{1 \over 4} R - {1 \over 4} F_{M N} F^{M N} - {3 \over 2L^2} \right]~.
}
We work in units where $L=1$.  This action has equations of motion
\eqn\Bb{\eqalign{
W_{MN}&\equiv R_{MN} + 3 g_{MN} - 2 F_{M P} F_{N}^{~P} + \half g_{MN} F_{PQ}F^{PQ}=0 \cr
Y^N&\equiv \nabla_M F^{MN} = 0~.}}
The dyonic black brane solution to \Bb\ is given by
\eqn\Bc{\eqalign{
ds^2 &= 2 dv dr - U(r) dv^2 + r^2dx^i dx^i \cr F & = {\rho \over r^2} dr \wedge dv +B dx^1 \wedge dx^2
}}
where U(r) is the function
\eqn\Bd{
U(r)= r^2 + {\rho^2 + B^2 \over r^2} - {2\epsilon \over  r}~,
}
and $\rho, \eps$ and $B$ are unspecified constants which will later be identified with the charge and energy density of our dual fluid as well as an external magnetic field applied to the fluid. The black brane has a singularity at $r=0$ which is shielded by a horizon at $r=a$, defined by the largest real root of $U(a) = 0$.   We are using Eddington-Finkelstein coordinates $(r,v,x^1, x^2)$, which are smooth across this future horizon.\foot{These are related to Schwarzschild type coordinates via
$v= t+r_*(r)$ with ${dr_* \over dr}= {1\over U(r)}$.   This gives $ds^2 = -U(r)dt^2 + {dr^2 \over U(r)}+r^2 dx^i dx^i$.}  These coordinates do not cover the past horizon, and we correspondingly do not demand smoothness there. This is physically sensible in that we do not expect solutions of dissipative fluid dynamics to be well behaved when extended arbitrarily far back in time.

The Hawking temperature of the brane is
\eqn\Be{
T = {3 a \over 4 \pi} - {B^2 + \rho^2 \over 4 \pi a^3}~.
}
$\eps$ and $\rho$ are restricted to values such that $T\geq 0$.
The chemical potential can be read off from the Euclidean black hole solution in terms of the
the asymptotic value of $A_v$.  Recall that to obtain a smooth gauge field on the Euclidean section
we must choose a gauge such that $A_v$ vanishes at the horizon (no such requirement exists for the Lorentzian solution due to the different topology).  This fixes the asymptotic value of $A_v$ and gives $\mu = {\rho \over a}$.

Some further conventions:
Latin  indices $M,N,...$ run over all four spacetime  coordinates, while Greek indices $\mu, \nu, ...$
run over the three coordinates $(v, x^1,x^2)$.   Since $v$ plays the role of time on the boundary,
we will sometimes use  $v=x^0$.   The boundary theory will always see a Minkowski metric,
$\tilde{\gamma}_{\mu\nu}dx^\mu dx^\nu = -(dx^0)^2+dx^i dx^i$.    Indices on the boundary stress
tensor and currents are raised and lowered with this metric.

\subsec{Stress tensor and current}

The action \Ba\ should be supplemented with the boundary terms \refs{\HenningsonGX,\BalasubramanianRE}
\eqn\Bf{
S_{bndy} =  - {1 \over \kappa_4^2} \int_{\p M} d^3 x \sqrt{-\gamma}~ \theta - {2 \over \kappa_4^2} \int_{\p M} d^3 x \sqrt{-\gamma}~.
}
Here $\gamma$ is the boundary metric and $\theta = \gamma^{\mu\nu}\theta_{\mu\nu}$, where
$\theta_{\mu\nu}= - {1\over 2} ( \del_{\mu} n_{\nu} + \del_{\nu} n_{\mu})$ is the extrinsic curvature of the boundary, defined in terms of the outward pointing unit normal vector $n$.

The conformal boundary metric is defined as $\tilde{\gamma}_{\mu\nu} = \lim_{r\rightarrow \infty} {1\over r^2} \gamma_{\mu\nu}$.  Also, the boundary gauge field is defined as $\lim_{r\rightarrow \infty}A_\mu$, in a gauge where $n^M A_M=0$.    The boundary stress tensor and current are then defined as
\eqn\Bg{
\delta S = {1 \over \kappa_4^2} \int_{\p M} \sqrt{-\tilde{\gamma}} \left(2 J^{\mu} \delta A_{\mu} +
T^{\mu \nu} \delta \tilde{\gamma}_{\mu \nu}\right)~.
}
Explicitly \BalasubramanianRE,
\eqn\Bh{\eqalign{
J^{\mu} &= r^2 F^{\mu r} \cr
T^{\mu \nu} &= {r^5 \over 2} \left[ \theta^{\mu \nu} - \theta \gamma^{\mu \nu} - 2 \gamma^{\mu \nu} \right]~.
}}
Implicit in \Bh\ is the large $r$ limit, as well as a projection of $T^{\mu\nu}$ parallel to the boundary (since the orthogonal component does not appear in \Bg.)

Electromagnetic gauge invariance implies current conservation,
\eqn\Bi{ \nabla_\mu J^\mu =0~.}
Invariance under diffeomorphisms generated by vector fields tangent to the boundary yields the (non) conservation equation
\eqn\Bj{ \nabla_{\nu} T^{\mu \nu} = F^{\mu \nu} J_{\nu}~.}
Tracelessness of the stress tensor follows from invariance under diffeomorphisms shifting the radial location of the boundary
\eqn\Bk{ \tilde{\gamma}_{\mu\nu} T^{\mu\nu} =0~.}
In particular, the latter invariance follows from the absence of logarithmic divergences in the bulk action, the presence of which would necessitate adding a non-diff invariant counterterm \HenningsonGX.

Applied to the solution \Bc\ we find
\eqn\Bl{\eqalign{ T^{\mu\nu} &= {\rm diag} (\epsilon, {1\over 2} \epsilon, {1\over 2}\epsilon) \cr J^\mu &= (\rho , 0 ,0)~,}}
which demonstrates that $\eps$ and $\rho$ are in fact energy and charge densities.

\subsec{Gravitational derivative expansion}

The dyonic black brane supergravity solution of \Bc\ represents a fluid in global thermodynamic equilibrium.  We would like to modify this solution in order to account for long wavelength hydrodynamic fluctuations,\foot{Although we use the word ``fluctuations", the disturbances are allowed to have large amplitude so long as their wavelength is large. This is the sense in which we are doing {\it nonlinear} hydrodynamics.}  so we begin by considering an approximate solution which looks locally like the dyonic brane.  We start with \Bc\ but allow $\eps$, $B$ and $\rho$ to be slowly varying functions of the spacetime coordinates $x^{\mu}$:
\eqn\Bm{\eqalign{
\left[ds^{(0)}\right]^2 &= 2 dv dr - U(r, x^{\mu}) dv^2 + r^2dx^i dx^i \cr F^{(0)} & = {\rho(x^{\mu}) \over r^2} dr \wedge dv +B(x^{\mu}) dx^1 \wedge dx^2
}}
where $U(r, x^{\mu})$ is given by
\eqn\Bn{
U(r, x^{\mu})= r^2 + {\rho^2(x^{\mu}) + B^2(x^{\mu}) \over r^2} - {2\epsilon(x^{\mu}) \over  r}~.
}
This approximate solution represents a good starting point for two reasons.  First, in small neighborhoods, it approximates the true dyonic black brane.  In the second place, this approximate solution approaches an exact solution in the limit of vanishing derivatives along the spacetime coordinates.  If derivatives are small, then we ought to approach an exact solution to the supergravity equations of motion by solving these equations order by order in a derivative expansion, and this is exactly how we proceed.

But first, we deal with some technicalities in defining the gauge potential corresponding to the field strength in the second line of \Bm .  We might consider an $A^{(0)}$ that can be written in the form
\eqn\ba{
A^{(0)} = - {\rho(x^{\mu}) \over r} dt  + \half B(x^{\mu}) \eps_{ij} x^i \wedge dx^j + A^{E}_{\alpha}(x^{\mu}) dx^{\alpha},
}
where $A^{E}_{\alpha}(x^{\mu})$ is purely electric, and $E_i = \p_v A^E_i - \p_i A^E_v$.  For constant $\rho, B, \eps$ and $A^{E}_{\alpha}$ this form of $A^{(0)}$ reproduces the field strength of \Bm.  In addition, this form is generic about any point since shifts in the origin correspond to constant shifts in $A^{(0)}_{\alpha}$.

But there is also a problem with the expression \Bm, because generically derivatives of $B(x^{\mu})$ do not correspond to derivatives of the magnetic field $F_{12}^{(0)}$.  If we want to match up derivatives then we need to correct $A^{(0)}$ in \Bm\ order by order in a derivative expansion.  For example in order to match $\p F$ and $\p^2 F$  we need to add terms of the form
\eqn\bb{
\left[{1 \over 4}\p_y B y^2 - {1 \over 12}\p_y^2 B y^3\right]dx - \left[{1 \over 4} \p_x B x^2 - {1 \over 12}\p_x^2 B x^3\right]dy~.
}
We note that it is possible to add new terms to this expression order by order in the derivative expansion to match up our definition of B in $A^{(0)}$ with the magnetic field and its derivatives in the boundary.

We add corrections to \Bm\ order by order in a derivative expansion in order to find a solution to the equations \Bb. As explained in \NLFDFG, the equations of motion can be solved ``tubewise" by working in small neighborhoods near
a given $x^\mu$ location, say $x^\mu=0$.  Each of these tubes then corresponds to a small neighborhood of local thermodynamic equilibrium in the boundary fluid.  We expand the metric and gauge fields as
\eqn\Bo{\eqalign{ g & =g^{(0)}(\eps,\rho,B,E) + \ve g^{(1)}(\eps,\rho, B, E)+ \ve^2 g^{(2)}(\eps,\rho,B,E) + {\cal O}(\ve^3) \cr A & = A^{(0)}(\eps,\rho, B, E) + \ve A^{(1)}(\eps,\rho, B, E)+ \ve^2 A^{(2)}(\eps,\rho,B,E) + {\cal O}(\ve^3)~,}}
where $g^{(0)}$ and  $A^{(0)}$  represent the lowest order solution corresponding to the dyonic black brane with variable $\rho$, $\eps$, $B$, and $E$. The derivatives of these parameters (and $E$ itself) act as sources for the higher order derivative corrections.  We have also introduced in this equation a parameter $\ve$ which formally labels the order of a term in our derivative expansion.

\Bb\ does not admit solutions starting with arbitrary values of $B, \rho, \eps$ and $E$.  Instead, we find that these parameters must satisfy constraint equations, which we then interpret as the fluid equations of motion in the boundary theory.  In general, we expect these equations to be modified order by order in a derivative expansion.  To allow for this, we express the energy and charge densities by an expansion,
\eqn\Bp{ \epsilon = \eps^{(0)}(\ve x^\mu)+ \ve \eps^{(1)}(\ve x^\mu) + \cdots ~,\quad\rho = \rho^{(0)}(\ve x^\mu)+ \ve \rho^{(1)}(\ve x^\mu) + \cdots ~. }
$B$ and $E$ however are specified by Dirichlet boundary conditions, so we wouldn't expect for $B$ or $E$ to be corrected order by order in our derivative expansion.  Nevertheless, we will formally write $B$ and $E$ in a similar fashion, with an eye towards S-duality,
\eqn\Bq{ B = B^{(0)}(\ve x^\mu)+ \ve B^{(1)}(\ve x^\mu) + \cdots ~,\quad E = \ve E^{(1)}(\ve x^\mu)+ \ve^2 E^{(2)}(\ve x^\mu) + \cdots ~.}
In the standard formulation where $E$ and $B$ are specified by Dirichlet boundary conditions $E^{n+1} = B^n = 0$ for $n>0$.  Note that this expansion in powers of $\ve$ is slightly different than the labeling used in section 2 where we just counted the number of derivatives acting on $\eps$ and $\rho$.  If we are working in a small neighborhood around $x^{\mu} = 0$, then it is convenient to set $\eps^{n > 0}(0) = \rho^{n > 0}(0) = E^n(0) = 0$.

The zeroth order solution preserves $SO(2)$ rotational symmetry, and this can be used to classify
the corrections to the metric and gauge fields.  We choose a gauge for the metric and gauge field
\eqn\Br{ A_r =0~,\quad g_{rr}=0~,\quad {g^{(0)}}^{\mu\nu}g_{\mu\nu}^{(n>0)} =0 ~,}
and decompose the fluctuations according to their $SO(2)$ representations:
\eqn\Bs{ A^{(n)} = A_v^{(n)} dv+ A_i^{(n)} dx^i~,}
with $A_v^{(n)}$  an $SO(2)$ scalar and $A_i^{(n)}$ an $SO(2)$ vector.  For the metric we write
\eqn\Bt{(ds^2)^{(n)}={k^{(n)}\over r^2}dv^2-2h^{(n)}dvdr +r^2 h^{(n)} dx^i dx^i + 2j^{(n)}_idv dx^i
+r^2 \sigma_{ij}^{(n)} dx^i dx^j~,}
In this expansion, $k^{(n)}$ and $h^{(n)}$ are $SO(2)$ scalars;  $j^{(n)}_i$ is an $SO(2)$ vector; and
$\sigma_{ij}^{(n)}$ is an $SO(2)$ symmetric traceless tensor.

We impose large $r$ boundary conditions on the $n>0$ components given by:
\eqn\Bu{ \eqalign{ &A_v^{(n)} \sim {1\over r^2}~,\quad A_i^{(n)} \sim {1\over r}  \cr & k^{(n)} \sim r^0~ \quad h^{(n)} \sim {1\over r^4}~,\quad j^{(n)}_i  \sim {1\over r}~,\quad \sigma^{(n)}_{ij} \sim {1\over r^3}~.}}
These conditions follow from a combination of the asymptotic AdS boundary conditions along with
the freedom to redefine coordinates as well as the zeroth order solution, as in \VanRaamsdonkFP.  In addition to these large $r$ boundary conditions, we must also demand that our solution be smooth across the future horizon at $r=a$.  For linear perturbations,  this condition is equivalent to demanding the presence of purely ingoing modes at the future horizon \HorowitzJD.

If we now plug the expansions in \Bo, \Bs , \Bt, into \Bb\ we arrive at the dynamical equations for the metric and gauge field corrections.  The $n$th order metric coefficients are determined by the components of the Einstein equations $W^{(n)}_{MN}=0$ with $M, N \neq v$.  These equations can be organized as
\eqn\Bv{\eqalign{ W^{(n)}_{rr}& = -{1\over r^4}\p_r(r^4 \p_r h^{(n)}) - S^{(n)}_{(h)} =0
\cr
r^2 (UW_{rr})^{(n)}-W_{ii}^{(n)}& = \p_r \left(-{2\over r}k^{(n)}\right)  +\p_r \left(\p_r(r^2 U^{(0)}) h^{(n)}\right)-{8\over r^2}B^2 h^{(n)} +  4\rho^{(0)}\p_r A_v^{(n)}  - S^{(n)}_{(k)} =0
\cr
W^{(n)}_{ri}& ={1\over 2}r \p_r \left({1\over r^2}\p_r (r j_i^{(n)})\right) +{2\over r^2 } [\rho^{(0)}\delta_{ij}-B\epsilon_{ij}]\p_r A_j^{(n)}  - S^{(n)}_i =0
\cr
W^{(n)}_{ij}-{1\over 2}\delta_{ij} W^{(n)}_{kk}& =\p_r \left(-{1\over 2}r^2 U^{(0)}\p_r \sigma_{ij}^{(n)}\right) - S^{(n)}_{ij} =0
}}
The source terms denoted by $S$ are constructed from the solution at order $n-1$, and so are assumed
to be known.  Similarly,  two components of the Maxwell equations yield
\eqn\Bw{\eqalign{ Y^{(n)v} &= {1\over r^2} \p_r \left(-r^2 \p_r A_v^{(n)}-2\rho^{(0)}h^{(n)}\right) - V^{(n)}=0 \cr Y^{(n)i}&= {1\over r^2} \p_r\left( U^{(0)} \p_r A_i^{(n)} +{1\over r^2}[\rho^{(0)}\delta_{ij}+B\epsilon_{ij}]j^{(n)}_j  \right) - V^{(n)}_i=0  }}
with source terms $V$.
These equations are sufficient to determine the metric and gauge field corrections for arbitrary $\rho$, $\eps$, $B$, and $E$, and are referred to as the dynamical equations.  The remaining $Y^r$ and $W_{vM}$ are interpreted as constraint equations and will be shown to be equivalent to the nonlinear magnetohydrodynamic equations of motion up to second order in derivatives.

Once the metric and gauge field corrections are determined using \Bv\ and \Bw, the current and stress tensor can be found using  (a large $r$ limit is implicit):
\eqn\Bx{\eqalign{J^{v}& =\rho
\cr J^{i}& = -r^2\sum_n \p_r A^{(n)}_i - E_i \cr T^{vv} &=\eps
\cr T^{vi} &= -{3\over 4} r \sum_n j^{(n)}_i \cr T^{ij} &={1\over 2}\eps+
{3\over 4} r^3 \sum_n \sigma^{(n)}_{ij}   }}

\newsec{Solving the equations}

\subsec{Zeroth order solution}

At zeroth order we use the dyonic black brane solution with $\eps = \eps^{(0)}$, $\rho= \rho^{(0)}$, $B= B^{(0)}$, and $E = 0$.  By construction, this is a solution to the Einstein-Maxwell equations.  The current and stress tensor are
\eqn\Ca{\eqalign{ J^{\mu} &= (\rho,0,0) \cr  T^{\mu\nu}&= {\rm diag}(\eps,{1\over 2} \eps,{1\over 2} \eps)~.}}

\subsec{First order solution}

At first order we write
\eqn\Cb{\eqalign{
\eps(x^\mu)&=  \eps^{(0)} + \ve x^\mu \p_\mu \eps^{(0)}(0) \cr
\rho(x^\mu)& = \rho^{(0)} + \ve x^\mu \p_\mu \rho^{(0)}(0) \cr
B^{(0)}(x^{\mu}) &= B^{(0)} + \ve x^{\mu} \p_{\mu} B^{(0)}(0) \cr
A^{E~(0)}_{\alpha}(x^{\mu}) &= \ve x^{\mu} \p_\mu A^{E~(0)}_{\alpha}(0)~.
}}
The first order sources are built out of $\p_\mu \eps^{(0)}$, $\p_\mu \rho^{(0)},\p_{\mu} B^{(0)}$ and  $E^{(1)}$,  and read
\eqn\Cd{\eqalign{S^{(1)}_{(h)}& = S^{(1)}_{(k)} =  S^{(1)}_{i} =   S^{(1)}_{ij} =V^{(1)} = 0~,\cr   V^{(1)}_i& ={\p_i \rho^{(0)} + \eps_{ij} \p_j B^{(0)} \over r^4} = {\p_j M_{ij}^- \over r^4}
}}
All fluctuations with no sources are set to zero by the boundary conditions. The non-zero fluctuations which must be determined are $\p_r A^{(1)}_i$ and $j^{(1)}_i$.

Integrating $Y^{(1)i} = 0$ gives
\eqn\Ce{
U^{(0)} \p_r A_i^{(1)} + { M_{ij}^- j_j^{(1)} \over r^2} + {\p_j M_{ij}^- \over r} = - c_i^{(1)}~,
}
where $c_i^{(1)}$ is a constant of integration, independent of $r$.

Proceeding as in \NMFG\ it is easy to obtain
\eqn\Cf{\eqalign{
j^{(1)}_i(r) &= -U^{(0)}(r) M_{ij}^+ \int_{\alpha^{(1)}_j}^r \! dr'\left({\beta^{(1)}_j +{4\over r'}c_j^{(1)} +{2\over r'^2} {\p_j M_{ij}^- \over U^{(0)}(r')^2}} \right) \cr
\p_r A_i^{(1)}(r) &= - {M_{ij}^-j_j^{1}(r) + r \p_j M_{ij}^- + r^2 c_i^{(1)} \over r^2 U^{(0)}(r)} ~.}}
We fix the integration constants by imposing \Bu\ and demanding that the fluctuations be smooth across the future horizon at $r = a$.  This fixes $\alpha^{(1)}_j=\infty$ and
\eqn\Cg{
\beta^{(1)}_i = -{4 \over a^{(0)}}\left(1+ {U'^{(0)}(a^{(0)}) \over a^{(0)}U''^{(0)}(a^{(0)})}\right)c_i^{(1)}-{2 \over {a^{(0)}}^2}\left(1+ {2U'^{(0)}(a^{(0)}) \over a^{(0)}U''^{(0)}(a^{(0)})}\right) \p_j M_{ij}^-~. }
With this choice of integration constants we find that  $j^{(1)}_i$ has the large $r$ behavior
\eqn\Ch{ j^{(1)}_i(r)  = {M_{i j}^+ \beta^{(1)}_j \over 3r} +{\cal O}({1\over r^2})~.}
To find the current $J^i$ we need
\eqn\Ci{
r^2 F^{i r~(1)} = - M_{ij}^- j_j^{(1)} - E_i^{(1)} - {\p_i \rho^{(0)} \over r} - U^{(0)} \p_r A_i^{(1)}~. }
At large $r$, we use this expression and \Bx\ to find
\eqn\Cj{
J^i = c_i^{(1)} - E_i^{(1)} ~,
}
which tells us that
\eqn\Ck{
c_i^{(1)} = N_i^-~.
}

The stress tensor can be found at this order using \Bx
\eqn\Cl{\eqalign{
T^{vi} & =-{1\over 4} M_{ij}^+ \beta^{(1)}_j   \cr
[T^{ij}]^{st}&=0~. }}
where the symmetric traceless part of a matrix is defined according to
\eqn\Cla{
[M_{ij}]^{st} = {1\over 2}\left(M_{ij} +M_{ji} - \delta_{ij} M_{kk} \right)~.
}

If we now combine \Cg\ and \Cl\ with our definition of the transport coefficients \Ah , $T^{0i} = c_1 x^1_i + c_2 x^2_i + c_3 x^3_i + c_4 x^4_i$, then it is easy to derive \Ax\
\eqn\Ax{\eqalign{c_1 & = {3(a^4+M) \over 4aM} \cr
c_2 &=0 \cr
c_3 & = {3a^4 +M \over 4a^2 M}
\cr
c_4 &=0~.}}
As advertised, we find the surprising result that two of the transport coefficients vanish.

We interpret the remaining $W_{vM}$ and $Y^r$ equations as constraints on the allowed values of the fluid variables.  If we use our solutions for $j_i^{(1)}$ and $\p_r A_i^{(1)}$ found in equations \Ce\ and \Cf, then it can be shown that these equations reduce to
\eqn\Cm{
\p_\mu J^{\mu}=0~,\quad \p_\nu T^{\mu\nu} = F^{\mu\nu} J_\nu
}
at order $\ve$ in our derivative expansion.  In particular these constraint equations take the form
\eqn\Cn{\eqalign{
&\p_v \rho^{(0)} =\p_v \eps^{(0)} = 0 \cr
&J^i = -{1 \over 2B} \eps_{ij} \p_j \eps - {\rho \over B} \eps_{ij} E_j \cr
& c^{(1)}_i  = - {1 \over B^{(0)} }\eps_{ij}\left[ \half \p_j \eps^{(0)} + M_{jk}^- E_k^{(1)} \right]~.
}}

\subsec{Second order solution}

$\eps$, $B$, $\rho$ and $A^E_{\alpha}(x^{\mu})$ are now given by expanding to order $\ve^2$,
\eqn\Co{  \eqalign{
\eps(x^\mu)&=  \eps^{(0)}(0) + \ve x^\mu \p_\mu \eps^{(0)}(0)+ {1\over 2}\ve^2x^\mu x^\nu \p_\mu \p_\nu\eps^{(0)}(0)+ \ve^2 x^\mu \p_\mu \eps^{(1)}(0)  \cr
\rho(x^\mu)& = \rho^{(0)}(0) + \ve x^\mu \p_\mu \rho^{(0)}(0) + {1\over 2}\ve^2x^\mu x^\nu \p_\mu \p_\nu\rho^{(0)}(0)+ \ve^2 x^\mu \p_\mu \rho^{(1)}(0) \cr
B(x^\mu)& = B^{(0)}(0) + \ve x^\mu \p_\mu B^{(0)}(0) + {1\over 2}\ve^2x^\mu x^\nu \p_\mu \p_\nu B^{(0)}(0) + \ve^2 x^\mu \p_\mu B^{(1)}(0) \cr
A^E_{\alpha}(x^{\mu}) &= \ve x^\mu \p_\mu A^{E~(0)}_{\alpha}(0) + {1\over 2}\ve^2 x^\mu x^\nu \p_\mu \p_\nu A^{E~(0)}_{\alpha}(0) + \ve^2 x^\mu \p_\mu A^{E~(1)}_{\alpha}(0)~.
}}
The second order sources work out to be
\eqn\Cp{\eqalign{S^{(2)}_{(h)}& = {2\over r^2}(\p_r A_i^{(1)})^2 \cr
S^{(2)}_{(k)}& =2U^{(0)}(\p_r A^{(1)}_i)^2 + {4 \over r^2} \rho^{(0)}\p_r A^{(1)}_i j^{(1)}_i-{4 \over r^2}B\epsilon_{ij}\p_i A^{(1)}_j \cr &\quad  +{2\over r} \p_i j^{(1)}_i  +\p_r \p_i j^{(1)}_i  -{2\over r} j^{(1)}_i \p_r j^{(1)}_i -{1\over 2}(\p_r j^{(1)}_i)^2 \cr
S^{(2)}_{i}& = 0\cr
S^{(2)}_{ij}& = -\p_r [\p_i j^{(1)}_j]^{st} -{2\over r} [j^{(1)}_i \p_r j^{(1)}_j]^{st} +{2\over r^2} [j^{(1)}_i j^{(1)}_j]^{st}+{1\over 2}[\p_r j^{(1)}_i \p_r j^{(1)}_j]^{st} \cr
&+[\left(2U^{(0)}\p_r A^{(1)}_i + {4\over r}\p_i \rho^{(0)} + 4 E_i + {4\over r^2} B \epsilon_{ik}j^{(1)}_k \right)\p_r A^{(1)}_j]^{st} \cr
V^{(2)}& ={1\over r^2}\p_r \p_i A_i^{(1)}-{1\over r^2}\p_r(j^{(1)}_i \p_r A_i^{(1)}) \cr
V^{(2)}_i& =   0~.         }}

It is not hard to work out the solutions to these equations following \NMFG.
\eqn\Cq{\eqalign{
h^{(2)}(r) &= -2 \int_\infty^r \! {dr' \over  r'^4} \int_\infty^{r'} \! dr'' \left( r'' \p_r A_i^{(1)}(r'')\right)^2 \cr
\p_r A_v^{(2)}(r) &= {1\over r^2} \int_\infty^r \! dr' r'^2 X_1(r') \cr
k^{(2)}(r) &= {1\over 2}r \int_\infty^r \! dr' X_2(r') \cr
j^{(2)}_i(r) &= -U^{(0)}(r) M_{ij}^+ \int_{\infty}^r \! dr'{\beta^{(2)}_j +{4\over r'}c_j^{(2)}\over U^{(0)}(r')^2} \cr
\sigma_{ij}^{(2)}(r) &= -2 \int_\infty^r  {dr' \over r'^2 U^{(0)}(r')} \int_{a^{(0)}}^{r'} \! dr'' S^{(2)}_{ij}(r''),
}}
where
\eqn\Cr{\eqalign{
X_1(r) &= - {2 \rho \over r^2} \p_r h^{(2)} \cr
X_2(r) &= \p_r \left(\p_r(r^2 U^{(0)}) h^{(2)}\right)-{8\over r^2}B^2 h^{(2)} +  4\rho^{(0)}\p_r A_v^{(2)}  - S^{(2)}_{(k)}~.
}}
The scalar sector does not contribute to the currents and we will not study the scalar sector solutions in any greater detail.

In the vector sector, $\p_r A_i^{(2)}$ is determined by the  analog of \Ce,
\eqn\Cs{ U^{(0)} \p_r A_i^{(2)} +{1\over r^2}M_{ij}^- j^{(2)}_j = -c^{(2)}_i~.}
Imposing regularity at the horizon tells us that
\eqn\Ct{ \beta^{(2)}_i = -{4 \over a^{(0)}}\left(1+ {U'^{(0)}(a^{(0)}) \over a^{(0)}U''^{(0)}(a^{(0)})}\right)c_i^{(2)}~. }
$\beta^{(2)}_i$ and $c_i^{(2)}$ now contribute to the stress tensor and current precisely as in \Cj\ and \Cl.

Also, the stress tensor conservation equation now fixes \foot{In many instances in this paper, we have used simply $N_i^{\pm}$ and $M_{ij}^{\pm}$ to refer to $N_i^{(1) \pm}$ and $M_{ij}^{(0) \pm}$; the difference is higher order in the derivative expansion and so we have not been overly careful to distinguish between the two.  Strictly speaking, we have $M_{ij}^{\pm} = M_{ij}^{(0)\pm} + M_{ij}^{(1)\pm} + \cdots$, and it is implicit that we are to take the lowest order when these quantities are present in derivative expansions.}
\eqn\Cu{ c_i^{(2)} = {1 \over B^{(0)} }\eps_{ij}\left[ - \half \p_j \eps^{(1)} + M_{jk}^+ E_k^{(2)} \right] = J_i^{(2)} + E_i^{(2)} = N_i^{(2)-} ~.}
The vector components of the stress tensor and current at this order take the same form as at first order.

In the tensor sector, the solution in \Cq\ is straightforward, and leads to
\eqn\Cv{
[T^{ij}]^{st} ={1\over 2}\int_{a^{(0)}}^{\infty}\! dr ~S^{(2)}_{ij}(r)~.
}

In order to write down the transport coefficients at this order we want to make use of the symmetries to organize the stress tensor according to the $S$-duality representations.  We use the operators $K_{ij}^{\pm} = M_{ik}^{\pm}M_{kj}^{\pm}$ with eigenvalue $- 1$ under $S$.  Below we list all of the terms which appear in the symmetric traceless part of the stress tensor at second order in derivatives, there are of course other terms consistent with the symmetries, but we list only the ones which appear in the fluid from our gravity dual.
\eqn\Cw{\eqalign{
\left[T^{ij}\right]^{st} =& b_{--}^{NN} \left[K^+_{ik} N_k^- N_j^- \right]^{st} + b_{+-}^{NN} \left[N_i^+ N_j^- \right]^{st} \cr
&+ b_{--}^{NM} \left[K^+_{ik} N_k^- \p_l M_{jl}^- \right]^{st} + b_{-+}^{NM} \left[ N_i^- \p_l M_{jl}^+ \right]^{st} + b_{+-}^{NM} \left[ N_i^+ \p_l M_{jl}^- \right]^{st} \cr
&+ b_{--}^{MM} \left[ K^+_{ik} \p_m M_{km}^- \p_l M_{jl}^- \right]^{st} + b_{-+}^{MM} \left[ \p_m M_{im}^+ \p_l M_{jl}^- \right]^{st} \cr
& + b^N \left[ M^+_{ik} \p_i N_{j}^- \right]^{st} + b^M \left[M_{ik}^+ \p_k \p_l M_{jl}^- \right]^{st}
.
}}
To read off the expressions for the transport coefficients we use \Cv\ and the expressions above.  We begin with the two linear transport coefficients
\eqn\Cx{\eqalign{
b^N &= - {a^2 \over 2 M} \cr
b^M &= - {a \over 2 M}~.
}}
As written in \Cw\ it may appear that these are only half the coefficients consistent with the symmetries, but there are only two linear transport coefficients present at this order in an arbitrary $S$-invariant fluid.  We emphasize that there is no mysterious cancelation in \Cx, unlike what we found at first order.

The nonlinear coefficients are more difficult to obtain.  The calculation is straightforward but not especially illuminating, so we simply state our result:
\eqn\Cy{\eqalign{
b_{--}^{NN} &=  {1 \over 2 \pi M T} + \half \int_a^{\infty} dr \left( -{2 \over r} j_N j'_N + {2 \over r^2}j_N j_N + \half j'_N j'_N - {2 \over r^2} j_N A_N \right) \cr
b_{--}^{MN} &= {3(a^4 + M) \over 4 \pi a M^2 T} + \int_a^{\infty} dr \left( -{j_M j'_N \over r}  - {j_N j'_M \over r} + 2 {j_N j_M \over r^2} + \half j'_N j'_M - {j_N A_M + j_M A_N \over r^2}\right) \cr
b_{--}^{MM} &= {3 a^2 \over 8 \pi M^2 T} + \half \int_a^{\infty} dr \left(-{2 \over r}j_M j'_M + {2 \over r^2}j_M j_M + \half j'_M j'_M - {2 \over r^2} j_M A_M \right) \cr
b_{+-}^{NN} &= {1 \over 2 \pi T} + \int_a^{\infty} dr A_N \cr
b_{-+}^{NM} &= {1 \over 4 \pi a T} + \int_a^{\infty} dr {A_N \over r} \cr
b_{+-}^{NM} &= {1 \over 4 \pi a T} + \int_a^{\infty} dr A_M \cr
b_{-+}^{MM} &= {1 \over 8 \pi a^2 T} + \int_a^{\infty} {A_M \over r}~,
}}
where the prime notation indicates a derivative with respect to $r$ and we have written the vector correction to the metric and gauge field at first order \Cf\ as
\eqn\Cz{\eqalign{
j_i^{(1)}(r) &= j_N(r) M_{ij}^+ N_j^- + j_M(r) M_{ij}^+ \p_k M_{jk}^- \cr
\p_r A_i^{(1)}(r) &= A_N(r) N_i^- + A_M(r) \p_j M_{ij}^- ~,
}}
with
\eqn\CAa{\eqalign{
j_N(r) &= {4 U(r) \over a} \int_{\infty}^r dr' {1 + {U'(a) \over a U''(a)} - {a \over r'} \over U^2(r)} \cr
j_M(r) &= {2 U(r) \over a^2} \int_{\infty}^r dr' {1 + {2 U'(a) \over a U'(a)} - {a^2 \over r^2} \over U^2(r)} \cr
A_N(r) &= - {\rho^2 + B^2 \over U(r) r^2}j_N - {1 \over U(r)} \cr
A_M(r) &= - {\rho^2 + B^2 \over U(r) r^2}j_M - {1 \over U(r) r} ~.
}}
Unlike the linear second order transport coefficients, there appear to be fewer coefficients present in the second order nonlinear sector than are predicted by symmetry. Coefficients of terms like $[K_{ik}^+ N^+_k N^+_j]^{st}$ and $[K_{ik}^- N^+_k N^+_j]^{st}$ are seen to be zero, and it is not hard to trace through our calculation to see that the cancelations at first order are responsible for the vanishing of these terms.   While it is possible that some (or even all) of these missing coefficients might be explained through a rigorous analysis of the entropy current at second order, we do not yet have an understanding of why some representations of the $S$-duality appear to be favored at this order.

In terms of these currents, it can be shown that the remaining constraint equations at this order reduce to the equations of motion in our fluid \Aa.

\newsec{Conclusion}

Let us review what has been achieved.  We first considered the fluid dynamics of a general S-invariant fluid at first order in the derivative expansion, independent of gravity or the
AdS/CFT correspondence. Even at this first nontrivial order there are a large number of transport coefficients, since we allow for
arbitrarily varying energy and charge  densities, and arbitrarily varying  background electromagnetic fields.
By using the constraints of symmetry and positive entropy production, we
found the most general form of the transport coefficients, and found that they could be expressed in terms of an arbitrary real function, whose argument is the single $S$-invariant dimensionless combination of the fluid variables, along with one real constant.   This result
implies many nontrivial relations among the various transport coefficients.  In principle, it would be possible to extend this analysis out to second or higher order in the derivative expansion, although the number of terms proliferates rapidly.

We then turned to the gravitational description of fluid dynamics in terms of black branes in an
asymptotically AdS$_4$ geometry.  By solving the Einstein-Maxwell equations order by order in a
boundary derivative expansion we were able to compute all transport coefficients up to second order. At first order our results were in agreement with those expected from our general fluid analysis, and we obtained the specific forms of the free function and constant that appeared in that analysis. Interestingly, this yielded vanishing values for two of the four transport coefficients, working in a basis natural under S-duality.  At second order we obtained new results for both linear and nonlinear transport coefficients.

S-invariance is a property shared by any fluid that has a holographic description in terms of four
dimensional Einstein-Maxwell theory.  One can hardly resist wondering whether such a fluid exists in nature.  A signal that one had found such a fluid would be to verify the relations among transport coefficients that we derived in section 2.  An even more remarkable result would be to find a fluid with vanishing (or small) values for the two transport coefficients which vanished in our gravitational computation.   This would be a smoking gun for holography, since this vanishing does not appear to be fixed by any obvious symmetry or consistency condition.

\bigskip
\noindent {\bf Acknowledgments:} \medskip \noindent    Work of PK is supported in part by NSF grant PHY-07-57702.

\appendix{A}{$S$ and $T$ operations:  bulk versus boundary}

In section 2, we referred to the observation of \WittenYA\ that  4-dimensional electric-magnetic $SL(2,Z)$ duality in the bulk is mapped to the operations we defined in \Ab\ and \Abc.  The purpose of this appendix is to establish this.

We begin with the S-operation and consider the most general field strength allowed by our boundary conditions and choice of gauge.  Using \Bm, \ba, and \Bo, we are restricted to considering an $F$ of the form
\eqn\Da{
F = {\rho \over r^2} dr \wedge dv + B dx^1 \wedge dx^2 + E_i dv \wedge dx^i + \sum_n \p_{M} A^{(n)}_{N}(r) dx^{M} \wedge dx^{N}~,
}
where all of the elements above are assumed to be functions of the boundary coordinates $x^{\mu}$.
Using \Bx\ and the boundary conditions we can write this as
\eqn\Dc{\eqalign{
F = & \left({\rho \over r^2} + {\cal O}({1 \over r^3}) \right) dr \wedge dv \cr
& + \left(B + {\cal O}({1 \over r}) \right)dx^i \wedge dx^j \cr
& + \left( E_i +  {\cal O}({1 \over r}) \right) dv \wedge dx^i \cr
& + \left( - {E_i + J_i \over r^2} + +  {\cal O}({1 \over r^3}) \right) dr \wedge dx^i
}}
Given the allowed form for the metric, as outlined in section 3.3, we compute the dual field strength
\eqn\Dd{\eqalign{
\star F = & \left({B \over r^2} + {\cal O}({1 \over r^3}) \right) dr \wedge dv \cr
& + \left(- \rho + {\cal O}({1 \over r}) \right)dx^1 \wedge dx^2 \cr
& + \left( - \eps_{ij}J^j +  {\cal O}({1 \over r}) \right) dv \wedge dx^i \cr
& + \left( \eps_{ij}{E^j + J^j \over r^2} + +  {\cal O}({1 \over r^3}) \right) dr \wedge dx^i~.
}}
Comparison of \Dc\ and \Dd\ demonstrates that electric-magnetic duality in the bulk exchanges the boundary CFT parameters according to \Ab ,
\eqn\Dda{\eqalign{\rho &\rightarrow B \cr
B &\rightarrow -\rho \cr
E_i  &\rightarrow -\epsilon_{ij}J^j \cr
J^i &\rightarrow -\epsilon^{ij}E_j  ~.  }}

Turning now to the  $T$ operation,  we consider adding to the action \Ba\ a $\theta$ term of the form
\eqn\De{
S_{\theta} = {\theta \over 8 \pi \kappa_4^2} \int \eps_{MNOP} F^{MN} F^{OP}~.
}
According to \Bg\ this term modifies our definition of the current to
\eqn\Df{
J^{\mu} = r^2 \left( F^{r \mu} + {\theta \over 4 \pi} \eps^{r \mu \nu \rho} F_{\nu \rho} \right)~,
}
where a large $r$ limit is implicit.  In the bulk, the $T$ operation corresponds to $\theta \rightarrow \theta + 2 \pi$.  In the boundary theory, this corresponds to \Abc ,
\eqn\Abc{\eqalign{ \rho &\rightarrow \rho +B \cr
B &\rightarrow B \cr
E_i  &\rightarrow E_i \cr
J^i &\rightarrow J^i -\epsilon^{ij}E_j ~.  }}

This establishes the relation between the $S$ and $T$ operations in the bulk and boundary.  It
also shows the different status of the two operations:  $S$ corresponds to a symmetry of the bulk
equations of motion, while $T$ corresponds to a change in the action of the theory.

\listrefs
\end